\documentstyle[epsfig,12pt]{article}
\newcommand{\be}{\begin{equation}}
\newcommand{\ee}{\end{equation}}
\newcommand{\ba}{\begin{eqnarray}}
\newcommand{\ea}{\end{eqnarray}}
\newcommand{\fslash}[1]{\mbox{$\!\not\!#1$}}
\newcommand{\vecp}[1]{\vec{#1}^{\,\, '}}

\begin{document}

\title{{\bf Faddeev calculation of pentaquark $\Theta^+$ \\in the Nambu-Jona-Lasinio
model-based\\ diquark picture}}

\author{
H. Mineo $^{a,b,c,1}$, J.A. Tjon $^{a,d}$, K. Tsushima $^{e,f,g,2}$ and
Shin Nan Yang $^{a}$\\
$^a$ Department of Physics,
National Taiwan University,
Taipei 10617, Taiwan\\
$^b$ Institute of Atomic and Molecular Sciences,
Academia Sinica, \\P.O.Box 23-166, Taipei 10617,
Taiwan\\
$^c$ Institute of Applied Mechanics, 
National Taiwan University, \\
Taipei 10617, Taiwan\\
$^d$ KVI, University of Groningen, The Netherlands\\
$^e$ National Center for Theoretical Sciences,
Taipei, Taipei 10617, Taiwan\\
$^f$ Grupo de F\'{\i}sica Nuclear and IUFFyM, Universidad de
Salamanca,\\ E-37008 Salamanca, Spain\\
$^g$ Thomas Jefferson National Accelerator Facility,
Theory Center, \\
Mail Stop 12H2, 12000 Jefferson Avenue,
Newport News, VA 23606, USA\\}

\date{\today}
\maketitle 
\abstract{ A Bethe-Salpeter-Faddeev (BSF) calculation is
performed for the pentaquark $\Theta^+$ in the diquark picture of
Jaffe and Wilczek in which $\Theta^+$ is a diquark-diquark-${\bar
s}$ three-body system.
 Nambu-Jona-Lasinio (NJL)
model is used to calculate the lowest order diagrams in the two-body
scatterings of ${\bar s}D$ and $D D$. With the use of coupling
constants determined from the meson sector, we find that ${\bar s}D$
interaction is attractive in $s$-wave while $DD$ interaction is
repulsive in $p$-wave. With only the lowest three-body channel
considered, we do not find a bound $ \frac 12^+$ pentaquark state.
Instead, a bound pentaquark $\Theta^+$ with $ \frac 12^-$ is
obtained with a unphysically strong vector mesonic coupling
constants.}

\footnotetext[1]{E-mail address: mineo@gate.sinica.edu.tw}
\footnotetext[2]{E-mail address: tsushima@jlab.org}
\newpage
\section{Introduction}
The report of the observation of a very narrow peak in the $K^+n$
invariant mass distribution \cite{LEPS03,CLAS03} around 1540 MeV
in 2003, a pentaquark predicted in a chiral soliton model
\cite{Diakonov}, triggered considerable excitement in the hadronic
physics community. It has been labeled as $\Theta^+$ and included
by the PDG in 2004 \cite{PDG04} under exotic baryons and rated
with three stars. Very intensive research efforts, both
theoretically and experimentally, ensued.

On the experimental side, practically all studies conducted after
the first sightings were confirmed by several other groups produced
null results, casting doubt on the existence of the five-quark state
\cite{BELLE04,HERA04}. Subsequently, PDG in 2006 reduced the rating
from three to one stars \cite{PDG04}.
More recently,  the ZEUS experiment at HERA \cite{HERA07} observed a
signal for $\Theta^+$ in  a high energy reaction, while H1
\cite{HERA07}, SPHINX \cite{SPHINX07} and CLAS \cite{CLAS05} did not
see it. This disagreement between the LEPS \cite{LEPS03} and other
experiments could possibly originate from their differences of
experimental setups and kinematical conditions. So the experimental
situation is presently not completely  settled
\cite{BELLE05,Hicks07,Chekanov07}.

Many theoretical approaches have been employed, in addition to the
chiral soliton model \cite{Diakonov}, including quark models
\cite{Stancu03}, QCD sum rules \cite{Kojo06}, and lattice QCD
\cite{Csikor03} to understand the properties and structure of
$\Theta^+$. Several interesting ideas were also proposed on the
pentaquark production mechanism. Review of the theoretical
activities in the last couple of years can be found in Refs.
\cite{Oka04,Jaffe05}.

One of the most intriguing theoretical ideas suggested for
$\Theta^+$ is the diquark picture of Jaffe and Wilczek (JW)
\cite{Jaffe03} in which $\Theta^+$ is considered as a three-body
system consisted of two scalar, isoscalar, color $\bf {\bar 3}$
diquarks ($D$'s) and a strange antiquark $(\bar s)$. It is based, in
part, on group theoretical consideration. It would hence be
desirable to examine such a scheme from a more dynamical
perspective.

The idea of diquark is not new. It is a strongly correlated quark
pair and has been advocated by a number of QCD theory groups since
60's \cite{Kobayashi66,Jaffe76,Anselmino93}. It is known that
diquark arises naturally from an effective quark theory in the low
energy region, the Nambu-Jona-Lasinio (NJL) model
\cite{NJL,Callan}. NJL model conveniently incorporates one of the
most important features of QCD, namely, chiral symmetry and its
spontaneously breaking which dictates the hadronic physics at low
energy. Models based on NJL type of Lagrangians have been very
successful in describing the low energy meson physics
\cite{Klevansky92,Takizawa90}. Based on relativistic Faddeev
equation the NJL model has also been applied to the baryon systems
\cite{Huang94,Ishii95}.  It has been shown that, using the
quark-diquark approximation, one can explain the nucleon static
properties reasonably well \cite{Asami95,Buck92}. If one further
take the static quark exchange kernel approximation, the Faddeev
equation can be solved analytically. The resulting forward parton
distribution functions \cite{Mineo99} successfully reproduce the
qualitative features of the empirical valence quark distribution.
The model has also been used to study the generalized parton
distributions of the nucleon \cite{Mineo}.
Consequently, we will employ NJL model to describe the dynamics of
a diquark-diquark-antiquark system. To describe such a
three-particle system, it is necessary to  resort to Faddeev
formalism.

Since the NJL model is a covariant-field theoretical model, it is
important to use relativistic equations to describe both the
three-particle and its two-particle subsystems. To this end, we
will adopt Bethe-Salpeter-Faddeev (BSF) equation \cite{Rupp88} in
our study. For practical purposes, Blankenbecler-Sugar (BbS)
\cite{BbS} reduction scheme will be followed to reduce the
four-dimensional integral equation into three-dimensional ones.

In Sec II, NJL model in flavor $SU(3)$ will be introduced with
focus on the diquark. The NJL model is then used to investigate
the antiquark-diquark and diquark-diquark interaction with
Bethe-Salpeter equation in Sec. III. In Sec. IV, we introduce the
Bethe-Salpeter-Faddeev equation and solve it for the system of
strange antiquark-diquark-diquark with the interaction obtained in
Sec. III. Results and discussions are presented in Sec. V, and we
summarize in Sec. VI.

\section{$\bf SU(3)_f$ NJL model and the diquark}

The flavor $SU(3)_f$ NJL Lagrangian takes the form \be {\cal
L}={\bar \psi}(i\fslash{\partial}-m)\psi +{\cal L}_I, \ee where
$\psi^T=(u,d,s)$ is the SU(3) quark field, and
$m=diag(m_u,m_d,m_s)$ is the current quark mass matrix.  ${\cal
L}_I$ is a chirally symmetric four-fermi contact interaction. By a
Fierz transformation, we can rewrite ${\cal L}_I$ into a Fierz
symmetric form ${\cal L}_{I,q{\bar q}} =\frac12({\cal L}_I+{\cal
F} ({\cal L}_I))$, where ${\cal F}$ stands for the Fierz
rearrangement. It has the advantage that the direct and exchange
terms give identical contribution.

In the $q{\bar q}$ channel, the chiral invariant ${\cal
L}_{I,q{\bar q}}$, is given by \cite{Klimt}

\ba
{\cal L}_{I,q{\bar q}}
&=&
G_1\left[
({\bar \psi}\lambda^a_f \psi)^2
-({\bar \psi}\gamma^5 \lambda^a_f \psi)^2
\right]
-G_2 \left[
({\bar \psi} \gamma^{\mu}\lambda^a_f\psi)^2
+({\bar \psi}\gamma^{\mu} \gamma^5\lambda^a_f \psi)^2
\right]\nonumber\\
&-&G_{3} \left[
({\bar \psi}\gamma^{\mu}\lambda^0_f  \psi)^2
+({\bar \psi}\gamma^{\mu}\gamma^5 \lambda^0_f \psi)^2
\right]
-G_{4} \left[
({\bar \psi}\gamma^{\mu}\lambda^0_f \psi)^2
-({\bar \psi}\gamma^{\mu}\gamma^5 \lambda^0_f \psi)^2
\right]\nonumber\\
&+&\cdots , \label{NJLLagrangian} \ea where $a=0\sim 8$, and
$\lambda_f^0=\sqrt{\frac23}I$. If we define $G_5$ by $-G_5({\bar
\psi}_i \gamma^{\mu}\psi_j)^2 =-(G_2+G_3+G_4) ({\bar \psi}_i
\gamma^{\mu}\lambda^0_f\psi_j)^2 -G_2 ({\bar \psi}_i
\gamma^{\mu}\lambda^8_f\psi_j)^2 $ where $i,j=u,d$, then
$G_3,G_4,G_5$ are related by $G_5=G_2+\frac23G_v,$ with $G_v\equiv
G_3+G_4.$ In passing, we mention that the conventionally used
$G_\omega$ and $G_\rho$ are related to $G_5, G_v$ by $G_\omega
=2G_5$ and $G_\rho=2G_5-\frac 43 G_v$.

For the diquark channel we rewrite ${\cal L}_I$ into an form
$({\bar \psi}A{\bar \psi}^T)(\psi^T B\psi)$, where $A$ and $B$ are
totally antisymmetric matrices in Dirac, isospin and color
indices. We will restrict ourselves to scalar, isoscalar diquark
with color and flavor in $\bf \bar 3$ as considered in the JW
model. The interaction Lagrangian for the scalar-isoscalar diquark
channel \cite{Vogl,Ishii} is given by
\be {\cal L}_{I,s} = G_s \left[ {\bar \psi}(\gamma^5 C )
\lambda_f^2 \beta^A_c {\bar \psi}^{T}\right] \left[ \psi^T (C^{-1}
\gamma^5 )\lambda_f^2
 \beta^{A}_c\psi \right],
\label{L_Is} \ee where $\beta^A_c=\sqrt{\frac32}\lambda^A
(A=2,5,7)$ corresponds to one of the color ${\bar 3}_c$ states.
$C=i\gamma^0\gamma^2$ is the charge conjugation operator, and
$\lambda's$ are the Gell-Mann matrices.

The Bethe-Salpeter (BS) equation for the scalar diquark channel
\cite{Vogl,Ishii} is given by \be \tau_s(q)=4iG_s -
2iG_s\int\frac{d^4k}{(2\pi)^4}
tr[(C^{-1}\gamma^5\tau_f^2\beta^A))S(k+q)
(\gamma^5C\tau_f^2\beta^A)S^T(-q)]\tau_s(q), \label{diquarktmatrix}
\ee where the factors $4$ and $2$ arise from Wick contractions.
$S(k)=(\fslash{k}-M+i\epsilon)^{-1}$ with $M\equiv M_u=M_d$, the
constituent quark mass of u and d quarks, generated by solving
the gap equation.
$\tau_s(q)$ is the reduced t-matrix which is related to
the t-matrix by $t_s(q)=(\gamma^5C\tau_f^2\beta^A_c)
\tau_s(q)(C^{-1}\gamma^5\tau_f^2\beta^A_c)$. The solution to Eq.
(\ref{diquarktmatrix}) is \be
\tau_s(q)=\frac{4iG_s}{1+2G_s\Pi_s(q^2)}, \label{taus} \ee with
\be \Pi_s(q^2)=6i\int\frac{d^4k}{(2\pi)^4}
tr_D[\gamma^5S(q)\gamma^5S(k+q)]. \label{pis} \ee

The gap equation for u, d and s quarks are given by \be
M_i=m_i-8G_1<{\bar q}_iq_i>, \label{gap} \ee with \be <{\bar
q}_iq_i>\equiv -iN_c\int\frac{d^4k}{(2\pi)^4} tr_D(S(k)),
\label{condense} \ee where $i=u,d,s$.

The loop integrals in Eqs. (\ref{pis}) and (\ref{condense})
diverge and we need to regularize the four-momentum integral by
adopting some cutoff scheme. With  regularization, we can solve
the gap equation and t-matrix of the diquark in Eqs.
(\ref{taus}) and (\ref{condense})  to determine the constituent
quark and diquark masses. However, since our purpose in this work
is not an exact quantitative analysis but rather a qualitatively
study of the interactions inside $\Theta^+$, we will not adopt any
regularization scheme and simply
use the empirical values of the constituent quark masses
$M=M_{u,d}=400$ MeV, $M_s=600$ MeV, and the diquark mass $M_D=600$
MeV as obtained in the study of the nucleon properties
\cite{Huang94,Ishii95,Asami95,Mineo99,Mineo}.

\section{Two-body interactions for strange antiquark-diquark $(\bf\bar
sD)$ and diquark-diquark ($DD$) channels}

In the JW model for $\Theta^+$, the two scalar-isoscalar, color
$\bf \bar 3$ diquarks must be in a color $\bf 3$ in order to
combine with $\bar s$ into a color singlet. Since $\bf 3$ is the
antisymmetric part of $\bf \bar 3 \times \bar 3 = 3 \oplus \bar
6$, the diquark-diquark wave function must be antisymmetric with
respect to the rest of its labels. For two identical
scalar-isoscalar diquarks $[ud]_0$, only spatial labels remain so
that the spatial wave function must be antisymmetric under space
exchange and the lowest possible state is $p$-state. Since in JW's
scheme, $\Theta^+$ has the quantum number of $J^P={\frac 12}^+$,
$\bar s$ would be in relative $s$-wave to the $DD$ pair.
Accordingly, we will consider only the configurations where $\bar
sD$ and $DD$ are in relative $s$- and $p$-waves, respectively.

 We will employ Bethe-Salpeter-Faddeev equation \cite{Rupp88} to
describe such a three-particle system of $\bar sDD$. For
consistency, we will use Bethe-Salpeter equation to describe
two-particles subsystems like $\bar sD$ and $DD$, which reads as,
 \be
 T = B + BG_0T,\label{BSEq}\ee
 where B is the sum of all two-body irreducible diagrams and $G_0$
 is the free two-body propagator. In momentum space, the resulting
 Bethe-Salpeter equation can be written as
 \be
 T(k',k;P)=B(k',k;P)+\int d^4k^{''}B(k',k^{''};P)G_0(k^{''};P)T(k^{''},k;P),
 \label{BSeq}\ee
 where $G_0$ is the free two-particle propagator in the intermediate states.
 $k$  and $P$ are, respectively, the relative and total momentum of the
 system.

 In practical applications, B is commonly
 approximated by the lowest order diagrams prescribed by the model
 Lagrangian and will be denoted by V hereafter. In
 addition, it is often to further reduce the dimensionality of the
 integral equation (\ref{BSeq}) from four to three, while preserving
 the relativistic two-particle unitarity cut in the physical region.
 It is well known (for example, Ref.
 \cite{Hung01}) that such a procedure is rather
 arbitrary and we will adopt, in this work, the widely employed
 Blankenbecler-Sugar (BbS) reduction scheme \cite{BbS} which,
 for the case of two spinless particles, amounts to replacing $G_0$ in
 Eq. (\ref{BSeq}) by
\ba G_0(k ,P)&=&\frac{1}{(P/2+k )^2-m_1^2}
\frac{1}{(P/2-k )^2-m_2^2}\nonumber\\
&\rightarrow& -i(2\pi)^4 \frac{1}{(2\pi)^3} \int
\frac{ds'}{s-s' +i\epsilon}\nonumber\\
&\times& \delta^{(+)}\left((P'/2+k )^2-m_1^2\right)
\delta^{(+)}\left((P'/2-k)^2-m_2^2
\right)\nonumber\\
&=& -2\pi i \delta\left(k_0
-\frac{E_1(|\vec{k}|)-E_2(|\vec{k}|)}{2}\right)
G^{BbS}(|\vec{k}|,s), \label{BbSthree1} \ea with \be
G^{BbS}(|\vec{k}|,s)=
\frac{E_1(|\vec{k}|)+E_2(|\vec{k}|)}
{2E_1(|\vec{k}|)E_2(|\vec{k}|)}
\frac{1}{s-(E_1(|\vec{k}|)+E_2(|\vec{k}|))^2 +i\epsilon},
\label{BbSthree2} \ee where $s=P^2$ and $P'=\sqrt{s'/s}P$. The superscript (+) associated with
the delta functions mean that only the positive energy part is
kept in the propagator, and $E_{1,2}(|\vec{k}|) \equiv
\sqrt{\vec{k}^2+m_{1,2}^2 }$.

\subsection{$\bf{\bar s}$D potential and the t-matrix}

In Fig. 1 we show the lowest order diagram, i.e., first order in
${\cal L}_{I,q{\bar q}}$ in ${\bar s}D$ scattering. Due to the
trace properties for Dirac matrices, only the scalar-isovector
$({\bar \psi}\lambda^a_f \psi)^2$, the vector-isoscalar $({\bar
\psi} \gamma^{\mu}\lambda^0_f\psi)^2$, and the vector-isovector
$({\bar \psi} \gamma^{\mu}\lambda^a_f \psi)^2$ will contribute to
the vertex $\Gamma$.
\begin{figure}[hbtp]
\begin{center}
\epsfig{file=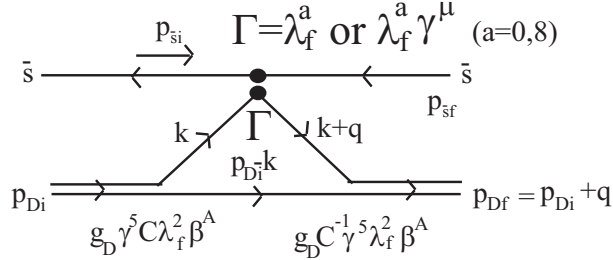,width=8cm}
\end{center}
\caption{${\bar s}$D potential
of the lowest order in ${\cal L}_{I,q{\bar q}}$.}
\end{figure}
Furthermore, the isovector vertex $(\bar{\psi}\Gamma\lambda_f^a
\psi)^2$ will not contribute since the trace in flavor space
vanishes, $\sum_{a=0}^8 (\lambda_f^a)_{33}tr_f(\lambda_f^2
\lambda_f^a\lambda_f^2)=0$. Thus only the vector-isoscalar term,
$({\bar \psi} \gamma^{\mu}\lambda^0_f\psi)^2$, remains.

For the on-shell diquarks, the lower part of Fig. 1 which
corresponds to the scalar diquark form factor, can be calculated
 as \ba
 (p_{Di}+p_{Df})^{\mu}F_v(q^2)
&=& i \int \frac{d^4 k}{(2\pi)^4} tr[ (g_D C^{-1}\gamma^5
\lambda_f^2\beta^A_c ) S(k+q)\gamma^{\mu}S(k) (g_D \gamma^5 C
\lambda_f^2\beta^A_c )
S^T(k-p_{Di})]\nonumber\\
&=& 6ig_D^2 \int \frac{d^4 k}{(2\pi)^4}
tr[S(k+q)\gamma^{\mu}S(k)S(p_{Di}-k)], \ea where we have made use
of the relations $C^{-1}(\gamma^{\mu})^T C=-\gamma^{\mu}$,
$tr_c[\beta^A_c \beta^A_c]=3$. $g_D$ is defined by \be
g_D^{-2}=-\left.\frac{\partial \Pi_D (p^2)} {\partial
p^2}\right|_{p^2=M_D^2},
 \ee with \be \Pi_D(p^2)\equiv 6i \int
\frac{d^4 k}{(2\pi)^4} tr[S(k)S(p-k)], \ee and $M_D$ is the
diquark mass. $F_v(0)$ is normalized as $2p^{\mu} F_v(0)=-g_D^2
\frac{\partial \Pi_D(p^2)}{\partial p_{\mu}}$, such that
$F_v(0)=1$. \footnote{In the actual calculation we use the dipole
form factor, $F_v(q^2)\equiv (1-q^2/\Lambda^2)^{-2}$ with
$\Lambda=0.84$ GeV since the $q^2$ dependence for $F_v(q^2)$ in
the NJL model is not well reproduced.}

Then the matrix element of the potential $V_{\bar sD}$  can be
expressed as \ba <{\bar s}_fD_f| V|{\bar s}_iD_i> &=&(-{\bar
v}(p_{{\bar s}i}))(-i V_{{\bar s}D})
(p_{Di},p_{Df})v(p_{{\bar s}f})\nonumber\\
&=&
(+16i) (-G_v)
(-{\bar v}(p_{{\bar s}i})) \gamma_{\mu}
v(p_{{\bar s}f})
\left[
(\lambda^0_f)_{33} \cdot tr_f
\left( \lambda^0_f (\lambda_f^2)^2
\right)
\right]\nonumber\\
&\times&(p_{Di}+p_{Df})^{\mu}\frac{F_v(q^2)}
{tr_f((\lambda_f^2)^2)}, \label{VsbarD} \ea i.e., \be {V}_{{\bar
s}D}=\frac{64}{3} G_v F_v(q^2) \tilde{V}_{{\bar
s}D}(p_{Di},p_{Df}), \label{VsbD} \ee with \be \tilde{V}_{{\bar
s}D}({p}_{Di},{p}_{Df}) =(\fslash{p}_{Di}+\fslash{p}_{Df})/2.
\label{tildeVsbD}\ee

Here the factor $+16i$ in Eq. (\ref{VsbarD}) arises from the Wick
contractions, and the factor $tr_f((\lambda_f^2)^2)$ in Eq.
(\ref{VsbarD}) is introduced to divide $F_v(q^2)$,
since the factor
$tr_f((\lambda_f^2)^2)$ is already included in
the expression of $F_v(q^2)$
by a trace in flavor $SU(3)_f$ space.

\begin{figure}[hbtp]
\begin{center}
\epsfig{file=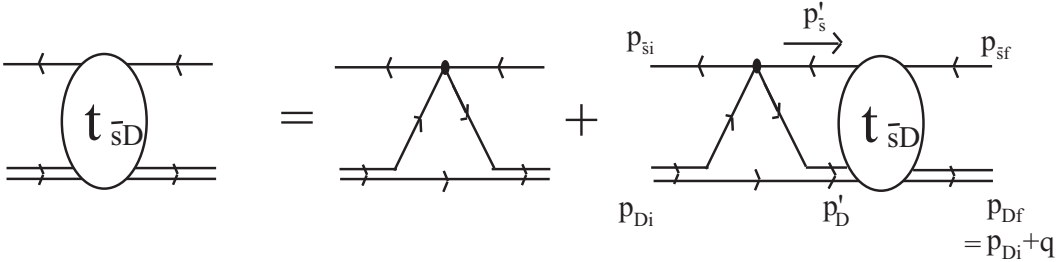,width=14cm}
\end{center}
\caption{The BS equation for ${\bar s}$D.}
\end{figure}

The three-dimensional scattering  equation for the ${\bar s}D$ system is now given by \ba
t_{{\bar s}D}(p_{Di},p_{Df})&=&
V_{{\bar s}D}(p_{Di},p_{Df})\nonumber\\
&+& 4\pi\int\frac{d |\vecp{p}_D| |\vecp{p}_D|^2}{(2\pi)^3} \frac12
\int_{-1}^1 dx_i G_{{\bar s}D}^{BbS}(|\vecp{p}_D|,s_2) K_{{\bar
s}D}(|\vec{p}_{Di}|,|\vecp{p}_D|,x_i)
t_{{\bar s}D}(\vecp{p}_D ,p_{Df}),\nonumber\\
\label{tsbD}
\ea
where $x_i\equiv \hat{p}_{Di}
\cdot \hat{p}_D^{\,\,'}$,
$\hat{p}\equiv\vec{p}/|p|$,
$s_2=(p_{Di}+p_{{\bar s}i})^2=(p_{Df}+p_{{\bar s}f})^2
$,
$p_{Di}^0=\sqrt{\vec{p}_{Di}^{\,\,2}+M_D^2}$,
$p_{Df}^0=\sqrt{\vec{p}_{Df}^{\,\,2}+M_D^2}$
and
\ba
K_{{\bar s}D}(|\vec{p}_{Di}|,|\vecp{p}_D|,x_i)
&\equiv& \frac{64}{3} G_v F_v((p'_D-p_{Di})^2)
\tilde{K}_{{\bar s}D}(p_{Di},p'_D)|_{{p'_{D}}^0
=\sqrt{ {\vec{p}}^{\,\,'2}_D+M_D^2}},\nonumber\\
\tilde{K}_{{\bar s}D}(p_{Di},p'_D) &=&
(\fslash{p}_{Di}+\fslash{p}_D^{\,\, '}) (-\fslash{p}_{\bar
s}^{\,\, '}+M_s)/2,\nonumber \ea with $M_s$ being the constituent
quark mass of ${\bar s}$ and $s$.

We also present the results for the interactions between
diquark and $\bar u$ or $\bar d$, which would be of interest when
we study non-strange pentaquarks. One can just repeat the
derivations we describe in the above and easily obtain \be
{V}_{{\bar u}D}={V}_{{\bar d}D}=\\
-16G_1 F_s(q^2)+32  G_5 F_v(q^2) \tilde{V}_{{\bar
s}D}(p_{Di},p_{Df}), \label{VubD} \ee in analogous to Eqs.
(\ref{VsbD}) and (\ref{tildeVsbD}).

We add in passing that, within tree approximation, the sign of the
potential for $sD$ is opposite to that of $V_{{\bar s}D}$ due to
charge conjugation, i.e., \be V_{sD}(p_{Df},p_{Di})=-V_{{\bar
s}D}(p_{Di},p_{Df}).\ee 
We can immediately write down the
scattering equation for the $sD$ as, \ba t_{sD}( p_{Df},p_{Di})
&=&V_{sD}( p_{Df},p_{Di})\nonumber\\
&+&
4\pi\int\frac{d |\vecp{p}_D| |\vecp{p}_D|^2}{(2\pi)^3}
\frac12 \int_{-1}^1 dx_f
G_{sD}^{BbS}(|\vecp{p}_D|,s_2)
K_{sD}(|\vec{p}_{Df}|,|\vecp{p}_D|,x_f)
t_{sD}(\vecp{p}_D ,p_{Di}),\nonumber\\
\label{tsD}
\ea
where $x_f\equiv \hat{p}_{Df}
\cdot \hat{p}_D^{\,\,'}$,
$G_{sD}^{BbS}(|\vecp{p}_D|,s_2)=G_{{\bar s}D}^{BbS}(|\vecp{p}_D|,s_2)$, and
\ba
K_{sD}(|\vec{p}_{Df}|,|\vecp{p}_D|,x_f)&\equiv &
\frac{64}{3} G_v F_v((p'_D-p_{Df})^2)
\tilde{K}_{sD}(p_{Df},p'_D)|_{{p'_{D}}^0
=\sqrt{ {\vec{p}}^{\,\,'2}_D+M_D^2}},\nonumber\\
\tilde{K}_{sD}(p_{Df},p'_D)&=&
- (\fslash{p}_{Df}+\fslash{p}_D^{\,\, '})
(\fslash{p}_s^{\,\, '}+M_s)/2,
\ea
with $p'_s=p'_{\bar s}$.

\subsection{Representation in $\rho$-spin notation}

In the ${\bar s}D$ (or $sD$)
center of mass system
the wave function which describes the relative motion
in $J=\frac12$, is given by the Dirac spinor of
the following form (see \cite{Tjon,Oettel}),

\ba
\Psi_{sD,m_s} (p_s^0,\vec{p}_s)
&=&
\left(
\begin{array}{c}
\phi_{s1}(p_s^0,|\vec{p}_s|)\\
\vec{\sigma} \cdot \hat{p}_s
\,\phi_{s2}(p_s^0,|\vec{p}_s|)\\
\end{array}
\right)\chi_{m_s},\\
\label{PsisD}
\Psi_{{\bar s}D,m_s} (p_{\bar s}^0,\vec{p}_{\bar s})
&=&
\left(
\begin{array}{c}
\vec{\sigma} \cdot \hat{p}_{\bar s}\,
\phi_{{\bar s}2}(p_{\bar s}^0,|\vec{p}_{\bar s}|)\\
\phi_{{\bar s}1}(p_{\bar s}^0,|\vec{p}_{\bar s}|)\\
\end{array}
\right)\chi_{m_s}, \nonumber\\
&=&
\gamma^5
\left(
\begin{array}{c}
\phi_{{\bar s}1}(p_{\bar s}^0,|\vec{p}_{\bar s}|)\\
\vec{\sigma} \cdot \hat{p}_{\bar s}
\,\phi_{{\bar s}2}(p_{\bar s}^0,|\vec{p}_{\bar s}|)\\
\end{array}
\right)\chi_{m_s},
\label{PsibarsD}\\
{\bar \Psi}_{sD}(p_s^0,\vec{p}_s)&\equiv&
\Psi_{sD}^{\dagger}(p_s^0,\vec{p}_s)\gamma^0,
\\
\label{barPsisD}
{\bar \Psi}_{{\bar s}D}(p_{\bar s}^0,\vec{p}_{\bar s})&\equiv&
\Psi_{{\bar s}D}^{\dagger}(p_{\bar s}^0,\vec{p}_{\bar s})\gamma^0,
\label{barPsibarsD}
\ea
where $\vec{p}_D=-\vec{p}_s=-\vec{p}_{\bar s}$,
i.e., $\Psi_{sD} (p_s^0,\vec{p}_s)
=\Psi_{sD} (p_s^0,-\vec{p}_D)$ and
$\Psi_{{\bar s}D} (p_{\bar s}^0,\vec{p}_{\bar s})
=\Psi_{{\bar s}D} (p_{\bar s}^0,-\vec{p}_D)$.
In the following we simply write
$p'_Q=|\vecp{p}_Q|, p'_{Qi(f)}=|\vecp{p}_{Qi(f)}|$
, $Q=s, {\bar s}$ or $D$.
Note that the index 1 (2) corresponds to large
(small) components for both ${\bar s}$ and $s$
quark spinors.

For a discretization in spinor space,
we define the complete set of $\rho$-spin notation
(\cite{Tjon,Gammel})
for the operators
${\cal O}_{sD}=V_{sD},t_{sD},\tilde{V}_{sD}$ and $
{\cal K}_{sD}=K_{sD},\tilde{K}_{sD}$
of $sD$:
\ba
{\cal O}_{sD,nm} (p_{Df},p_{Di}) &\equiv &
tr[ \Omega^{\dagger}_n (p_{sf})
{\cal O}_{sD}(p_{Df},p_{Di})\Omega_m(p_{si})],
\label{sDnm1}
\\
{\cal K}_{sD,nm} (p_{Df},p_D',x_f) &\equiv &
tr[ \Omega^{\dagger}_n (p_{sf})
{\cal K}_{sD}(p_{Df},p_D',x_f) \Omega_m(p'_s)],
\label{sDnm2}
\ea
where $n,m=1,2$,
$\Omega_1(p)=\frac{\Omega}{\sqrt{2}}$ and
$\Omega_2(p)=\vec{\gamma}\cdot \hat{p}
\frac{\Omega}{\sqrt{2}}$,
$\Omega=\frac{1+\gamma_0}{2}$.
$\Omega_1(p)$ and $\Omega_2(p)$ satisfy
$tr[ \Omega^{\dagger}_n (p) \Omega_m(p')]=
\delta_{n1}\delta_{m1}+
\hat{p}\cdot\hat{p}^{\,\, '}\delta_{n2}\delta_{m2}
$.

Concerning the ${\bar s}D$ spinor, the large and small components
can be reversed by $\gamma^5$, with the minus sign which comes
from the definitions Eqs. (\ref{PsibarsD}) and
(\ref{barPsibarsD}): ${\bar \Psi}_{{\bar s}D}{\cal O}\Psi_{{\bar
s}D}= -{\bar \Psi}_{sD}\gamma^5{\cal O}\gamma^5 \Psi_{sD}$. Then
we can define $\rho$-spin notation for ${\bar s}D$ i.e., ${\cal
O}_{{\bar s}D}=V_{{\bar s}D},t_{{\bar s}D}, \tilde{V}_{{\bar s}D}$
and ${\cal K}_{{\bar s}D}=K_{{\bar s}D} ,\tilde{K}_{{\bar s}D}$,
\ba {\cal O}_{{\bar s}D,nm} (p_{Di},p_{Df})
 &\equiv&
-tr[ \Omega^{\dagger}_n (p_{{\bar s}i})
\gamma^5 {\cal O}_{{\bar s}D}(p_{Di}
,p_{Df})\gamma^5 \Omega_m(p_{{\bar s}f})],
\label{sbDnm1}\\
{\cal K}_{{\bar s}D,nm} (p_{Di},p_D',x_i) &\equiv &
-tr[ \Omega^{\dagger}_n (p_{{\bar s}i})
\gamma^5 {\cal K}_{{\bar s}D}(p_{Di},p_D',x_i)
\gamma^5\Omega_m(p_{\bar s}')].
\label{sbDnm2}
\ea

From Eqs. (\ref{tsbD},\ref{tsD},\ref{sDnm1}-\ref{sbDnm2}), each
component $n$ $(n=1,2)$ of spinors for the ${\bar s}$D satisfy the
following quadratic equation: \ba && \phi^{\dagger}_{{\bar s}n}
(p_{{\bar s}i}) t_{{\bar s}D,nm}(p_{Di},p_{Df}) \phi_{{\bar s}m}
(p_{{\bar s}f}) =\phi^{\dagger}_{{\bar s}n} (p_{{\bar s}i}) \Bigl[
V_{{\bar s}D,nm}(p_{Di},p_{Df})\nonumber\\
&&+4\pi\sum_{l=1}^2
\int\frac{dp_D'} {(2\pi)^3}
p_D^{'2} \frac12\int_{-1}^1 dx_i
G_{{\bar s}D}^{BbS}(p_D',s_2)
K_{{\bar s}D,nl}(p_{Di},p_D',x_i)
t_{{\bar s}D,lm}
(p_D',p_{Df})
\Bigr] \phi_{{\bar s}m} (p_{{\bar s} f}).\nonumber\\
\label{tsbDnm}
\ea

A similar equation can be obtained for the $sD$ by exchanging
$i\leftrightarrow f$ and $s\leftrightarrow {\bar s}$ in Eq.
(\ref{tsbDnm}).

The explicit expressions of the $\rho$-spin notation
for
$\tilde{V}_{{\bar s}(s)D}$ and
$\tilde{K}_{{\bar s}(s)D}$
are given in appendix B.
We note that there are important relations:
\ba
V_{{\bar s}D,nm}(p,q)
&=& -V_{sD,nm}(p,q),\nonumber\\
V_{{\bar s}D}(p,q)&=& -V_{sD}(p,q),\nonumber\\
K_{{\bar s} D,nm} (|\vec{p}^{\,}|,|\vec{q\,}|,x_{pq})
&=& -K_{sD,nm}(|\vec{p}^{\,}|,|\vec{q}^{\,}|,x_{pq})
,\nonumber\\
K_{{\bar s}D}(|\vec{p}^{\,}|,|\vec{q}^{\,}|,x_{pq})
&=& - K_{sD} (|\vec{p}^{\,}|,|\vec{q}^{\,}|,x_{pq}).
\nonumber
\ea

By the partial wave expansion in Eq. (\ref{PWE}) in appendix A,
the BS equation for $t_{{\bar s}D,nm}$ in Eq. (\ref{tsbDnm}) for
$s$-wave can be written as \be t_{{\bar s}D,nm}^{l_{{\bar
s}D}=0}(p_{Di},p_{Df}) =V_{{\bar s}D,nm}^{l_{{\bar
s}D}=0}(p_{Di},p_{Df}) +4\pi \int \frac{d p_D^{\,'}}{(2\pi)^3}
p_D^{\,'2} \sum_{l=1}^2 G_{{\bar s}D}^{BbS}(p_D^{\,'},s_2)
K^{l_{{\bar s}D}=0}_{{\bar s}D,nl}(p_{Di},p_D^{\,'}) t_{{\bar
s}D,lm}^{l_{{\bar s}D}=0}(p_D^{\,'} ,p_{Df}). \label{tsbDint} \ee

\subsection{$DD$ potential and t-matrix}

In the case of $DD$ interaction, the lowest order diagrams are
depicted in Figs. 3(a) and (b), with (a) the quark rearrangement
diagram and (b) of the first order in ${\cal L}_{I, q\bar q}$,
respectively.

\begin{figure}[hbtp]
\begin{center}
\epsfig{file=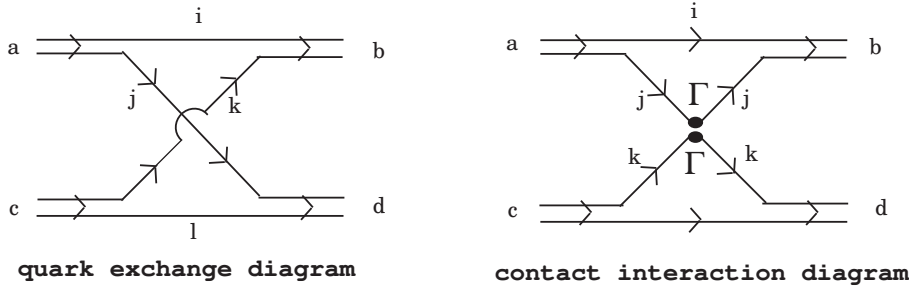,width=12cm}
\end{center}
\caption{Lowest order diagrams in $DD$ scattering.  }
\end{figure}

We first show that the quark exchange diagram in Fig. 3(a) does
not contribute due to its color structure, where $a\sim d$ and
$i\sim l$ denote the color indices of the diaquarks and quarks,
respectively. Since each diquark is in the color ${\bf\bar 3} $
\cite{Jaffe03,Vogl}, the color factor for the $qqD$ vertex is
proportional to $\epsilon_{aij}$.  Hence the color factor of the
quark exchange diagram is given by \be
\epsilon_{aij}\epsilon_{bik}\epsilon_{clk}\epsilon_{dlj}=
\delta_{ab}\delta_{cd}+\delta_{ad}\delta_{bc}. \label{quarkexh}
\ee As we discussed earlier, the color of the $DD$ pair inside
$\Theta^+$ is of ${\bf 3}$ in order to combine with ${\bar s}$ to form
a color singlet pentaquark. As color ${\bf 3}$ state is
antisymmetric under the exchange between diquarks in the initial
and final states, the matrix element of Eq. (\ref{quarkexh})
vanishes.

For the contact interaction diagram Fig. 3(b), only the direct
term is shown since the exchange term does not contribute as it
has the same color structure as the quark rearrangement diagram of
Fig 3(a). It is easy to see that the color structure of Fig.
3(b) is proportional to $\delta_{ab}\delta_{cd}$. Then the terms in
the interaction Lagrangian in Eq. (\ref{NJLLagrangian}) that can
give rise to non-vanishing contributions are: \be G_1({\bar
\psi}\lambda^a_f \psi)^2,\,\, -G_2({\bar
\psi}\gamma^{\mu}\lambda^a_f \psi)^2,\,\, -G_v ({\bar
\psi}\gamma^{\mu}\lambda^0_f \psi)^2, \ee with $a=0\sim 8$.

We next calculate the form factors, which diagrammatically
correspond to the lower part of diagram in Fig. 1. For
$\Gamma=\gamma^{\mu}\lambda_f^a$, we obtain \ba && {tr_f \left(
\lambda_f^a (\lambda_f^2)^2 \right) } (p_{Di}+p_{Df})^{\mu}
\frac{F_v(q^2)}{tr_f((\lambda_f^2)^2)}\nonumber\\
&=& \left( \sqrt{\frac23}\delta_{a0}+\sqrt{\frac13}\delta_{a8}
\right) (p_{Di}+p_{Df})^{\mu} F_v(q^2), \label{Fv} \ea and for
$\Gamma=\lambda_f^a$, we get \be {tr_f \left( \lambda_f^a
(\lambda_f^2)^2 \right)}
\frac{F_s(q^2)}{tr_f((\lambda_f^2)^2)}\nonumber\\
= \left( \sqrt{\frac23}\delta_{a0}+\sqrt{\frac13}\delta_{a8}
\right) F_s(q^2), \label{Fs} \ee where the factor
${tr_f((\lambda_f^2)^2)}$ in Eqs. (\ref{Fv}) and (\ref{Fs}) is
introduced by the same reason for Eq. (\ref{VsbarD}), and we have
used $tr(\lambda_f^2 \lambda_f^a \lambda_f^2)=
2(\sqrt{\frac23}\delta_{a0} +\sqrt{\frac13}\delta_{a8})$.

For the on-shell diquarks, $F_s(q^2)$ is calculated as\footnote{
Same as the case for ${\bar s}D$ potential, we use the dipole form
factor, $F_s(q^2)\equiv c_s (1-q^2/\Lambda^2)^{-2}$ with
$\Lambda=0.84$ GeV and $c_s$ is a constant. In the original NJL
model calculation with the Pauli-Villars (PV) cutoff, $c_s$ is
given by $F_s(0)=c_s=0.53$ GeV \cite{Mineo}.} \ba F_s(q^2) &=& i
\int \frac{d^4 k}{(2\pi)^4} tr[ (g_D C^{-1}\gamma^5
\lambda_f^2\beta^A ) S(k+q)S(k) (g_D \gamma^5 C \lambda_f^2\beta^A
)
S^T(k-p_{Di})]\nonumber\\
&=&
6ig_D^2 \int \frac{d^4 k}{(2\pi)^4}
tr[S(k+q)S(k)S(k-p_{Di})].
\ea

With the form factors $F_v(q^2)$ and $F_s(q^2)$ obtained in the
above, $V_{DD}$ is given by
\ba
-iV_{DD}(\vec{p}_{Di},\vec{p}_{Df}) &=& +128i \left[ G_1 F_s^2(q^2) -\left(
G_2+\frac23 G_v \right) (p_{D1i}+p_{D1f})\cdot
(p_{D2i}+p_{D2f})F_v^2(q^2)\right]\nonumber\\
&=&128i \left[ G_1 F_s^2(q^2)
-G_5 (p_{D1i}+p_{D1f})\cdot(p_{D2i}+p_{D2f})F_v^2(q^2)\right],
 \label{128i}
\ea
where the factor $+128i$ in a first line of
Eq. (\ref{128i}) comes from the Wick
contractions, and in a second line
we have used the relation between couplling constants
in meson sectors; $G_5= G_2+\frac23 G_v$ which is
explained in section 2.
The momenta of the diquarks in the initial and
final states in Fig. 4 are given by
\ba p_{D1i(f)}
&=&(\sqrt{s_2}/2,\vec{p}_{Di(f)})
,\nonumber\\
p_{D2i(f)} &=&(\sqrt{s_2}/2,-\vec{p}_{Di(f)}),
\ea with
$q=p_{D1f}-p_{D1i}=p_{D2i}-p_{D2f}$.
$s_2=4(\vec{p}_{Di}^{\,2}+M_D^2)=4(\vec{p}_{Df}^{\,2} +M_D^2)$ is
the $DD$ center of mass energy squared.

\begin{figure}[hbtp]
\begin{center}
\epsfig{file=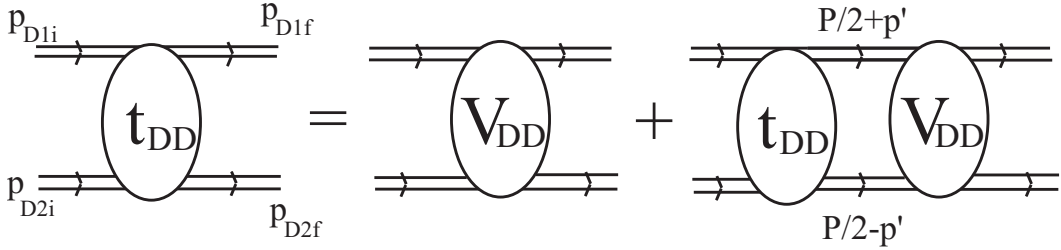,width=14cm}
\end{center}
\caption{BS equation for $DD$.}
\end{figure}

As in the case of $\bar sD$ scattering, we use the BbS
three-dimensional reduction scheme and the resulting equation for $DD$
scattering reads as\be
 t_{DD}(\vec{p}_{Df},\vec{p}_{Di})
= V_{DD}(\vec{p}_{Df},\vec{p}_{Di})
+ \int\frac{d^3 p'}{(2\pi)^3}
V_{DD}(\vec{p}_{Df},\vecp{p})
G_{DD}^{BbS}(|\vecp{p}|,s_2)
t_{DD}(\vecp{p},\vec{p}_{Di}),
\ee
with
\ba
G_{DD}^{BbS}(|\vecp{p}|,s_2)&=&
\frac{1}{
4E_D(|\vecp{p}|)(s_2/4-E_D(|\vecp{p}|)^2+i
\epsilon)}\nonumber\\
&=&\frac{1}{4E_D(|\vecp{p}|)(\vec{p}_{Df}^{\,\,2}
-\vec{p}^{\,\,'2}+i\epsilon)},
\label{GDD}
\ea
with $E_D(|\vec{p}^{\,\,'}|)
=\sqrt{\vec{p}^{\,\,'2}+M_D^2}$.

In the JW model for $\Theta^+$, the diquark-diaquark spatial wave
function must be antisymmetric and we will consider here only the
lowest configuration, namely, $DD$ are in relative $p$-wave. Partial
wave expansion of Eq. (\ref{PWE}) then gives \be t_{DD}^{l=1}
(p_f,p_i)=V_{DD}^{l=1} (p_f,p_i) +4\pi \int \frac{dp'
}{(2\pi)^3}p'^2 G_{DD}^{BbS}(p',s_2)
V_{DD}^{l=1}(p_f,p')t_{DD}^{l=1}(p',p_i),
\label{tDDlresultofs} \ee with $p_{i(f)}\equiv|\vec{p}_{Di(f)}|,
p'\equiv|\vecp{p}|$.

\section{Relativistic Faddeev equation}
\subsection{3-body Lippmann-Schwinger equation}

For a system of three particles with momenta $\vec k_i's \,
(i=1,2,3)$, we introduce the Jacobi momenta with particle 3 as a
special choice: \ba
\vec{k}_1 &=& \mu_1 \vec{P} + \vec{\tilde{p}} + \alpha_1\,\, \vec{\tilde{q}}_3\nonumber\\
\vec{k}_2 &=& \mu_2 \vec{P} - \vec{\tilde{p}} + \alpha_2\,\, \vec{\tilde{q}}_3\nonumber\\
\vec{k}_3 &=& \mu_3 \vec{P} + \alpha_3\,\, \vec{\tilde{q}}_3, \ea
with $\sum \mu_n =1$ and $\alpha_3=  -\alpha_1-\alpha_2$. For the
coefficients we find $\mu_n = m_n/M$,\,\, $M =m_1+m_2+m_3$,\,\, and
$\alpha_1= m_1/m_{12},\,\, \alpha_2=m_2/m_{12},\,\, \alpha_3=-1$,
where $m_{ij}=m_i+m_j\,\,(i\neq j).$
 In terms of
the Jacobi momenta the total kinetic energy is given by: \be
K_{tot} = \frac{P^2}{2 M} + \frac{\tilde{p}\,^2}{2 m_{12}}
+\frac{{\tilde{q}_3}^2}{2m_{(12)3}},\ee  where $m_{(ij)k}= {m_k
m_{ij}}/{M}$.

New integration variables are chosen to be: $\tilde{p} = f_{p3}\,\,
p$ with $f_{p3}=\sqrt{2 m_{12}}$ and $\tilde{q_3}=f_{q3}\,\,q$ with
$f_{q3}=\sqrt{2 m_{(12)3}}$, and in general for cyclic $(ijk)$,
$f_{pi}=\sqrt{2 m_{jk}}$ and $f_{qi}=\sqrt{2 m_{(jk)i}}$. In terms
of the new integration variables we have \be K_{tot}=\frac{P^2}{2 M}
+p^2+q^2, \ee and the 3-body Lippmann-Schwinger equation for the
T-matrix becomes: \be T(\vec{p},\vec{q})=V+ {f_{p3}}^3 {f_{q3}}^3
\int\,\,\frac{d^3p'}{(2\pi)^3}\int\,\, \frac{d^3q'}{(2\pi)^3}\,\,
V\,\, G_3(p',q')\,\,T(\vec{p}\,',\vec{q}\,'),\label{LS-T}\ee with
$G_3(p,q)= 1/(z-K_{tot})$. The parameter $z$ is implicit in the
arguments of $T$ and $G_3$ in Eq. (\ref{LS-T}), a convention to be
followed hereafter.

Similarly we define the Jacobi momenta $\vec{p}_i,\vec{q}_i$ with
particle $i$ as the special choice. The momenta are related to each
other as \ba \vec {p}_i = a_{ij} \vec {p}_j + b_{ij} \vec {q}_j,
\hspace{2.0cm} \vec {q}_i = c_{ij} \vec {p}_j + d_{ij} \vec {q}_j ,
\ea where $(ijk)$ are cyclic, and $a_{ij} = - [m_i m_j /(m_i + m_k
)(m_j + m_k )]^{1/2} $, $b_{ij} = \sqrt {1 - a_{ij}^2} = - b_{ji} $,
$c_{ij}=-b_{ij}$ and $d_{ij}=a_{ij}$.

It can be shown that the total angular momentum is
related to the angular momentum $\vec {l}_{pi}$ and $\vec {l}_{qi}$ by
\begin{equation}
\label{eq5}
\vec {L} = \sum\limits_{i = 1}^3 {\left( {\vec {r}_i \times \vec {k}_i }
\right)} = \sum\limits_{i = 1}^3 {\left( {\vec {l}_{pi} + \vec {l}_{qi} }
\right)} + \vec {l}_c .
\end{equation}

With these three choices of Jacobi momenta we may introduce
corresponding 3-particle states $|>_n$ where particle $n$ plays a
special role. For the 3-particle T-matrix we have \be
<\vec{k}_1,\vec{k}_2,\vec{k}_3|T|\alpha>=_n<\vec{p}_n,\vec{q}_n|T|\alpha>,
\ee or in terms of the Faddeev amplitudes $T_n$, \be
<\vec{k}_1,\vec{k}_2,\vec{k}_3|T|\alpha> = T_1(\vec{p}_1,
\vec{q}_1)+T_2(\vec{p}_2,\vec{q}_2)+T_3(\vec{p}_3,\vec{q}_3), \ee
with $T_n(\vec{p}_n,\vec{q}_n)=_n<\vec{p}_n,\vec{q}_n|T_n|\alpha>$.

For the pentaquark system we now chose particles 1 and 3 as the
diquark and particle 2 to be the ${\bar s}$. The Faddeev equations
for $T=T_1+T_2+T_3$ with $T_i = t_i + \sum\limits_{j \ne i} {t_i }
G_2 (s)T_j\,\,\,(i=1,2,3)$, with $t_i$ denoting the two-body
t-matrix between particle pair $(jk)$, become \ba
T_1(\vec{p}_1,\vec{q}_1) &=& {f_{p3}}^3 {f_{q3}}^3\int
\frac{d^3p'_3}{(2\pi)^3}\int \frac{d^3q'_3}{(2\pi)^3}\,\, K_{13}
\,\,G_3(p'_3,q'_3)\,\,T_3(\vec{p}_3\,',\vec{q}_3\,')\nonumber\\
&+& {f_{p2}}^3 {f_{q2}}^3 \int \frac{d^3p'_2}{(2\pi)^3} \int
\frac{d^3q'_2}{(2\pi)^3}\,\, K_{12}\,\,G_3(p'_2,q'_2)\,\,
T_2(\vec{p}_2\,',\vec{q}_2\,'),\label{T_1} \ea where the channels 1
and 3 correspond to $D({\bar s}D)$ states and channel 2 to the
${\bar s}(DD)$ states. Since diquarks obey Bose-Einstein statistics,
we have $T_3(\vec{p}_3,\vec{q}_3)=T_1(-\vec{p}_3,\vec{q}_3)$ and
$T_3(\vec{p}_3,\vec{q}_3)=T_1(-\vec{p}_1,\vec{q}_1)$. We note that
the symmetry property which requires the amplitude $T$  be
anti-symmetric with respect to interchange of the 2 diquarks is
automatically satisfied by the angular momentum content
$L=l_{q1}=l_{p2}=1,l_{p1}=l_{q2}=0$.

The ${\bar s}(DD)$ T-matrix $T_2$ satisfies \be
T_2(\vec{p}_2,\vec{q}_2) = 2 fp^3_1 fq^3_1\int
\frac{d^3p'_1}{(2\pi)^3} \int \frac{d^3q'_1}{(2\pi)^3} \,\, K_{21}
\,\,G_3({p}_1',{q}_1')\,\,T_1(\vec{p}_1\,',\vec{q}_1\,'). \label{T2}
\ee

The kernels $K_{13}$ and $K_{12}$ are expressed in terms of the
${\bar s}D$ t-matrix \be K_{13}=K_{12}= t_{{\bar
s}D}(\vec{p}_1,{\vec{p}_1}\,';z-q_1^2)\,\,
\frac{{(2\pi)^3}}{f_{q_1}\,^3} \delta^{(3)}[\vec{q}_1-\vec{q}_1
\,']. \label{K13} \ee

Similarly the kernel $K_{21}$ is given by
\be
K_{21}= t_{DD}(\vec{p}_2,\vec{p}_{2}\,';z-q_2^2)
\,\,\frac{{(2\pi)^3}}{f_{q_2}\,^3}
\delta^{(3)}[\vec{q}_2-\vec{q}_2\,'].
\label{K21}
\ee

The term with $K_{13}$ can be worked out by making use of the
$\delta$-function relation \be \delta ^{(3)}\left[ {\vec {q}_1 -
{\vec {q}}_1\,'} \right] = \frac{2}{q_1} \delta \left( {q_1^2 -
{{q}_1'}^2 } \right) \delta \left( {\cos \theta _{q_3 } - \cos
\theta_{{q}_3' } } \right)\delta \left( {\phi_{{q}_3' } - \phi_{q_3
} } \right), \ee and the linear relation $\vec{q}_1\,'=c_{13}
\vec{p}_3\,' +d_{13} \vec{q}_3\,'$, which lead to \ba \delta
^{(3)}\left[ {\vec {q}_1 - \vec{q}_1\,' } \right]
 &= &\frac{1}{q_1 c_{13} d_{13} {p}_3' {q}_3' }
\delta \left( {\cos \theta _{p_3' q_3' } -
\frac{q_1'^2 - c_{13}'^2 p_3'^2 - d_{13}^2 q_3'^2 }
{2c_{13} d_{13}p_3' q_3' }} \right)\nonumber\\
&\times&
\delta \left( {\cos \theta _{q_3 }
- \cos \theta_{q_3' } } \right)\delta
\left( {\phi_{q_3' } - \phi_{q_3 } } \right).
\label{deltafunc}
\ea
We mention that similar expression for a delta function
in the term $K_{12}$ can also be obtained by replacing
$3 \rightarrow 2$.

Performing a partial wave expansion for the $D({\bar s}D)$ amplitude
\be T_1(\vec{p}_1,\vec{q}_1) = 4 \pi Y_{lp_1 0}^*
(\Omega_{p_1})Y_{lq_1 0} (\Omega_{q_1}) T_1^L(p_1,q_1), \ee and for
the $\bar sD$ t-matrix $t_{\bar {s}D} (\vec {p}_1 ,\vec {p}_1\,' ;z
- q_1^2 )$, \be t_{\bar {s}D} (\vec {p}_1 ,\vec {p}_1 \,' ;z - q_1^2
) = 4\pi {Y_{l{p_1} 0}^* (\Omega_{p_1} )Y_{lp_1 0} (\Omega_{p'_1} )}
t_{\bar {s}D}^{(l_{p1} )} ({p}_1 ,{p}_1' ;z - q_1^2 ), \ee yield \ba
&&T_1^L(p_1,q_1) \nonumber\\
&& =   c_3 \int_0^\infty {q'_3}^2 dq'_3 \int_{A_{13}}^{B_{13}}
{p'_3}^2 dp'_3\,\, t^{(lp_1)}_{{\bar s}D}(p_1,p'_1;z-q_1^2)\,\,
X_{13}\,
\frac{1}{c_{13}\,d_{13}\,q_1\,p'_3\,q'_3}\,\,G_3(p'_3,q'_3)\,\,
T_3^L(p'_3,q'_3)\nonumber\\
&&+  c_2 \int_0^\infty {q'_2}^2 dq'_2 \int_{A_{12}}^{B_{12}}
{p'_2}^2 dp'_2 \,\, t_{{\bar s}D}^{(lp_1)}(p_1,p'_1;z-q_1^2)\,\,
X_{12}\, \frac{1}{c_{12}\,d_{12}\,q_1\,p'_2\,q'_2}
\,\,G_3(p'_2,q'_2)\,\,
T_2^L(p'_2,q'_2),\nonumber\\
\ea
with
\be
c_3= \frac{2}{\sqrt{ \pi}} ({f_{p3} f_{q3}}/{f_{q1}})^3,\,\,
c_2= \frac{2}{\sqrt{ \pi}} ({f_{p2} f_{q2}}/{f_{q1}})^3,
\ee
and where the boundaries $A,B$ for the $p'$ integration can easily be
found from the condition $q^2_1={q'_1}^2$ in Eq. (\ref{deltafunc}),
given by
\ba
A_{ij} &=& \left|\frac{ c_{ij} {q}'_j + q_i }{d_{ij} }\right|\\
B_{ij} &=& \left|\frac{ c_{ij} {q}'_j - q_i }{d_{ij} }\right|,
\label{ABbound}
\ea

For the ${\bar s}(DD)$ amplitude $T_2$, partical wave expansion
gives,  \ba T_2^L(p_2,q_2) &=& 2 c_1 \int_0^\infty {q'_1}^2 dq'_1
\int_{A_{21}}^{B_{21}} {p'_1}^2 dp'_1\,\,\nonumber\\
&\times&t_{DD}^{(lp_2)}(p_2,p'_2;z-q^2_2)\,\,
X_{21}\,\,\frac{1}{c_{21}\,d_{21}\,q_2\,p'_1\,q'_1}
\,\,G_3(p'_1,q'_1)\,\,T_1^L(p_1',q_1'),
\ea
where $A_{21}$ and $B_{21}$ are given by Eq. (\ref{ABbound}),
and
\be
c_1=\frac{2}{\sqrt{ \pi}} ({f_{p1} f_{q1}}/{f_{q2}})^3.
\ee

In the above equations
$X_{ij}$ are angular momentum functions depending on the states we
  consider. In our case, the ${\bar s}D$ 2-body channel is a s-wave,
$lp=0$, and the $DD$ channel a p-wave, $lp=1$.  Hence, for the 3-body
channel with total angular momentum $L=1$ we have for the $D({\bar s}D)$
3-body channnel $lp_1=0,lq_1=L$ and $lp_3=0,lq_3=L$, while for
${\bar s}(DD)$ $lp_2=1,lq_2=0$. The obtained $X_{ij}$ have the form
\be
X_{13}= \frac{1}{4 \pi \sqrt{3}}
Y_{lq_3 0}(\theta_{q_3\,q_1}), \,\,
X_{12}=\frac{1}{4 \pi\sqrt{3}}
Y_{lq_2 0}(\theta_{q_2\,q_1}),\,\,
X_{21}=\frac{1}{4\pi \sqrt{3}}
Y_{lp_2 0}(\theta_{p_2\,p_1} ).
\label{X}
\ee

\subsection{Relativistic Faddeev equations}

Following Amazadeh and Tjon \cite{Amazadeh} (see also \cite{Rupp88})
we adopt the relativistic quasi-potential prescription based on a
dispersion relation in the 2-particle subsystem. Then the 3-body
Bethe-Salpeter-Faddeev equations have essentially the same form as
the non relativistic version.Taking the representation with particle
3 as special choice we may write down for the 3-particle Green
function a dispersion relation of the (1,2)-system, i.e., \be
G_3(p_3,q_3;s_3) = \frac{E_1(k_1)+E_2(k_2)}{E_1(k_1)\,E_2(k_2)}\,
\frac{1}{s_3-q_3^2-(E_1(k_1)+E_2(k_2))^2}, \ee with
$E_1(k_1)=\sqrt{k_1^2+m_1^2},\,E_2(k_2)=\sqrt{k_2^2+m_2^2},$ and
$s_3=P^2$ being the invariant 3-particle energy square. In the
3-particle cm-system we have $\sqrt{s_3}=M+E_b$. The resulting
2-body Green function with invariant 2-body energy square $s_2$ has
then the form of the BSLT quasi-potential Green function \be
G_2(p_3;s_2) =
\frac{E_1(k_1)+E_2(k_2)}{E_1(k_1)\,E_2(k_2)}\,
\frac{1}{s_2-(E_1(k_1)+E_2(k_2))^2}.
\ee
 This quasi-potential
prescription for $G_3$ has obviously the advantage that the 2-body
t-matrix in the Faddeev kernel satisfies the same equation as the
one in the 2-particle Hilbert space with only a shift in the
invariant 2-body energy. So the structure of the resulting 3-body
equations are the same as in the non relativistic case.

\section{Results and discussions}

In the NJL model some cutoff scheme  must be adopted since the NJL
model is non-renormalizable. However, in this work we will not use
any cutoff scheme but simply employ the dipole form factors for
the scalar and vector vertices. Namely, the NJL model is only used to
study the Dirac, flavor and color structure of the ${\bar s}D$ and
$DD$ potentials.

For the values of the masses $M_{u,d}$, $M_s$ and $M_D$, we use
the empirical values $M=M_u=M_d=400$ MeV and $M_s=M_D=600$ MeV
\cite{Mineo}. We will treat the coupling constants $G_i$ $(i=1\sim
5)$ in Eq. (\ref{NJLLagrangian}) as free parameters. For the
${\bar s}D$ channel, it depends only on $G_v=G_3+G_4 =\frac32
(G_5-G_2)$ as seen in Eq. (\ref{VsbarD}).

In the NJL model calculation with the Pauli-Villars (PV) cutoff
regularization \cite{Mineo},
the coupling constants $G_{\pi}$, $G_{\rho}$ and
$G_{\omega}$ are related with the parameters used in our work by
$G_1=G_{\pi}/2$, $G_2=G_{\rho}/2$ and $G_5=G_{\omega}/2$.
Thus by using the values of mesonic coupling constants in the
NJL model, $G_v$ is determined as
$G_v=\frac32(G_{\omega}/2-G_{\rho}/2)
=\frac32(7.34/2-8.38/2)=-0.78$ GeV$^{-2}$.
We remark that the sign
of $G_v$ is definitely negative since experimentally omega meson is
heavier than the rho meson. Then the interaction between ${\bar
s}$ and diquark in $s$-wave is attractive, as can be seen from the
${\bar s}D\,\, s$-wave phaseshift shown in Fig. 5 with $G_v=-0.78$
GeV$^{-2}$, while the interaction between ${s}$ and diquark is repulsive
which can be seen in Fig. 6.
In both figures we find that the magnitudes of the phaseshift is
within 10 degrees, that is, $G_v=-0.78$ GeV$^{-2}$ gives
very weak interaction between ${\bar s}$ ($s$) and diquark.
As we can see in Figs. 5 and 6,
generally the phaseshift in $s$-wave is more sensitive
to three momentum than that in $p$-wave.
We note that ${\bar s}D$ and $sD$
phaseshift are not symmetric around the $p_E$ axis,
which can be understood from the decompositions of
$t_{sD}$ and $t_{{\bar s}D}$ in the spinor space
in appendix B.
We further mention that if $G_v$ is determined
from the $\Lambda$ hyperon mass
$M_{\Lambda}=1116$ MeV within the $sD$ picture, one obtains
$G_v=6.44$ GeV$^{-2}$,
which is different from $G_v=-0.78$ GeV$^{-2}$
determined from meson sector in the NJL model in sign.
In this case the rho meson mass is larger
than the omega meson mass, that is,
the vector meson masses are not correctly reproduced.

$DD$ phaseshift is plotted in Fig. 7
where we have used the values of coupling constants
$G_1=G_{\pi}/2=5.21$ GeV$^{-2}$ and $G_5=G_{\omega}/2=3.67$
GeV$^{-2}$ which are determined from meson sectors
in the NJL model calculation with the
Pauli-Villars cutoff \cite{Mineo}.
We can easily see that the phaseshift $\delta_l$ is definitely
negative i.e., the $DD$ interaction is repulsive,
and its dependence on three momentum $p_E$ is very strong
and almost proportional to $p_E$ both for $s$-wave and $p$-wave.
This strong $p_E$ dependence of phaseshift comes
from the $p_E^2$ dependence of a second term
$(p_{D1i}+p_{D1f})\cdot(p_{D2i}+p_{D2f})$
in Eq. (\ref{128i}).

The $G_v$ dependence of the ${\bar s}D$ binding energy, $E_{{\bar
s}D}$, is presented in Fig. 8. We find that the ${\bar s}D$ bound
state begins to appear around $G_v=-5\sim -6$ GeV$^{-2}$, becomes
more deeply bound as $G_v$ becomes more negative. It is easily seen
that $E_{{\bar s}D}$ is almost proportional to $G_v$. However even
for the case of a weakly bound state with $|E_{{\bar s}D}|$ less
than $0.1$ GeV, it will require a value of $-G_v=5\sim 6$ GeV$^{-2}$
which is about eight times larger than  the $-G_v$ determined from
meson sector in the original NJL model with the PV cutoff
regularization.

For the calculation of the pentaquark binding energy we use the
relativistic three-body Faddeev equation which is introduced in
section 4. If the pentaquark state is in $J^P=\frac12^+$ state
with which we are concerned in the present paper, the total force
is attactive but there is no pentaquark bound state.

On the other hand if the pentaquark state is in $J^P=\frac12^-$
state, a bound pentaquark state begins to appear when $G_v$
becomes more negative than $-8.0 \,$ GeV$^{-2}$, 
a value inconsistent with what is required to predict 
a bound $\Lambda$ hyperon with
$M_{\Lambda}=1116$ MeV in a quark-diquark model as mentioned in
Sec. 5. The lowest configuration which would correspond to a
$J^P=\frac12^-$ state is for the spectator $\bar s$ to be in
$p-$wave w.r.t. to a $DD$ pair in $p-$wave, or alternatively
speaking, the spectator diquark in relative $s$-wave to $\bar sD$
in $s$-wave. Our results for the binding energy of a $J^P=\frac 12^-$
pentaquark state for the case with and without $DD$ channel are
given in Table 1. It is found that although the $DD$ interaction
is repulsive, including the $DD$ channel gives an additional
binding energy which is leading to the more deeply pentaquark
boundstate. It is because the coupling to the $DD$ channel is
attractive due to the sign of the effective kernel $K_{21}$ in
Eqs. (\ref{T2}, \ref{K21}). This depends on the recoupling
coefficients $X_{21}$, $X_{12}$ in Eq. (\ref{X}) and the 2-body
t-matrices.

\begin{table}
\begin{center}
\begin{tabular}[htbp]{|c|c|c|}
\hline
$G_v [GeV^{-2}]$ & $E_B^0(5q) [MeV]$ & $E_B(5q) [MeV]$\\
\hline
-8.0 & 47  & 77\\
\hline
-9.0 & 87 & 139  \\
\hline
-10.0 & 132 & 205 \\
\hline
-12.0 & 226 & 333 \\
\hline
-14.0 & 316 & 505 \\
\hline
\end{tabular}
\caption{The binding energy of $J^P =\frac 12^{-}$ pentaquark
state. $E_B^0 (5q)$ ($E_B (5q)$) is the binding energy without
(including) the $DD$ channel.}
\end{center}
\end{table}

%


In Fig. 9 (10)
the phaseshift of ${\bar s}D$ is plotted,
where the coupling constant is fixed at $G_v=-8.0$ GeV$^{-2}$
($G_v=-14.0$ GeV$^{-2}$).
It is easily seen that in Figs. 9 and 10 the phaseshift of
${\bar s}D$ in $s$-wave is positive for small $p_E<0.3$ GeV
and $p_E<0.45$ GeV, but
it changes the sign around $p_E = 0.3$ and $p_E = 0.45$ GeV,
thus the phaseshift of ${\bar s}D$ in $s$-wave is
very sensitive to three momentum $p_E$.
Whereas the phaseshift of ${\bar s}D$ in $p$-wave is definitely
positive.

In Fig. 11 we plot the phaseshift of ${s}D$ with the coupling constant
$G_v=-14.0$ GeV$^{-2}$ which is same as the one used in
Fig. 10.
Different from the phaseshift of ${\bar s}D$
the phaseshifts of  $sD$ in $s$ and $p$-wave do not change the sign for higher three momentum $p_E$, i.e., the sign of the phaseshifts
are definitely negative.

From the above results we find that even if we use a very strong
coupling constant $G_v$ which is unphysical
because it gives much larger mass difference of
rho and omega mesons than the experimental
value, $M_{\omega}-M_{\rho}=13$ MeV,
it is impossible to obtain the pentaquark bound state with
$J^P=\frac12^+$.
With only the $J=\frac12$
three-body channels considered, we do not find a
bound $J^P=\frac12^+$ pentaquark state.
The $J^P=\frac12^-$ channel is more attractive,
resulting in a bound pentaquark state in this channel, but for
unphysically large values of vector mesonic coupling constants.



\section{Summary}

In this work, we have presented a Bethe-Salpeter-Faddeev (BSF)
calculation for the pentaquark $\Theta^+$ in the diquark picture
of Jaffe and Wilczek in which $\Theta^+$ is treated as a
diquark-diquark-${\bar s}$ three-body system. The
Blankenbecler-Sugar reduction scheme is used to reduce the
four-dimensional integral equation into three-dimensional ones.
The two-body diquark-diquark and diquark-${\bar s}$ interactions
are obtained from the lowest order diagrams prescribed by the
Nambu-Jona-Lasinio (NJL) model. The coupling constants in the NJL
model as determined from the meson sector are used. We find that
${\bar s}D$ interaction is attractive in $s$-wave while $DD$
interaction is repulsive in $p$-wave. Within the truncated
configuration where $DD$ and ${\bar s}D$ are restricted to $p$-
and $s$-waves, respectively, we do not find any bound $ \frac
12^+$ pentaquark state, even if we turn off the repulsive $DD$
interaction. It indicates that the attractive ${\bar s}D$
interaction is not strong enough to support a bound $DD\bar s$
system with $J^P=\frac 12^+$.

However, a bound pentaquark with $J^P=\frac 12^-$ begins to appear
if we change the vector mesonic coupling constant $G_v$ from
$-0.78$ GeV$^{-2}$, as determined from the mesonic sector, to
around $G_v=-8$ GeV$^{-2}$. And it becomes more deeply bound as
$G_v$ becomes more negative.

\section*{Acknowledgements}
This work was supported in part by the National Science Council of
ROC under grant no. NSC93-2112-M002-004 (H.M. and S.N.Y.). J.A.T.
wishes to acknowledge the financial support of NSC for a visiting
chair professorship at the Physics Department of NTU and the warm
hospitality he received throughout the visit. K.T. acknowledges
the support from the Spanish Ministry of Education and Science,
Reference Number: SAB2005-0059.

\begin{figure}[hbtp]
\begin{center}
\epsfig{file=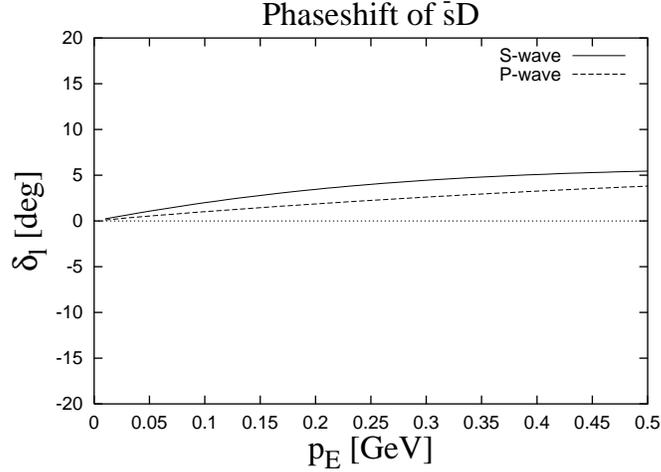,scale=0.5}
\end{center}
\caption{Three momentum $p_E$ dependence of the phaseshift
$\delta_l$ for the ${\bar s}D$ interaction with
the coupling constant $G_v=-0.78$ GeV$^{-2}$. }
\end{figure}

\begin{figure}[hbtp]
\begin{center}
\epsfig{file=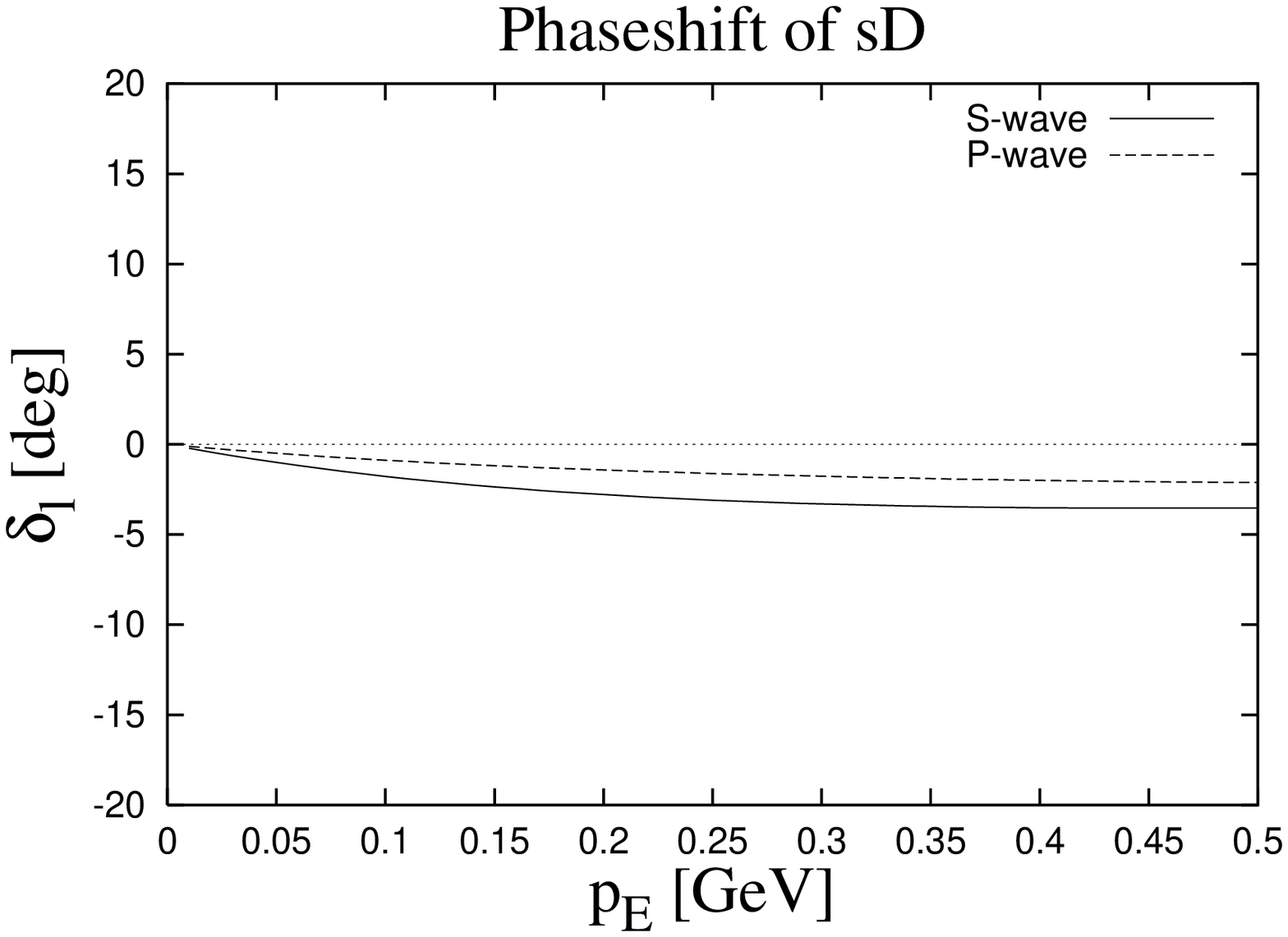,scale=0.5}
\end{center}
\caption{Three momentum $p_E$ dependence of the phaseshift
$\delta_l$ for the ${s}D$ interaction with
the coupling constant $G_v=-0.78$ GeV$^{-2}$. }
\end{figure}

\begin{figure}[hbtp]
\begin{center}
\epsfig{file=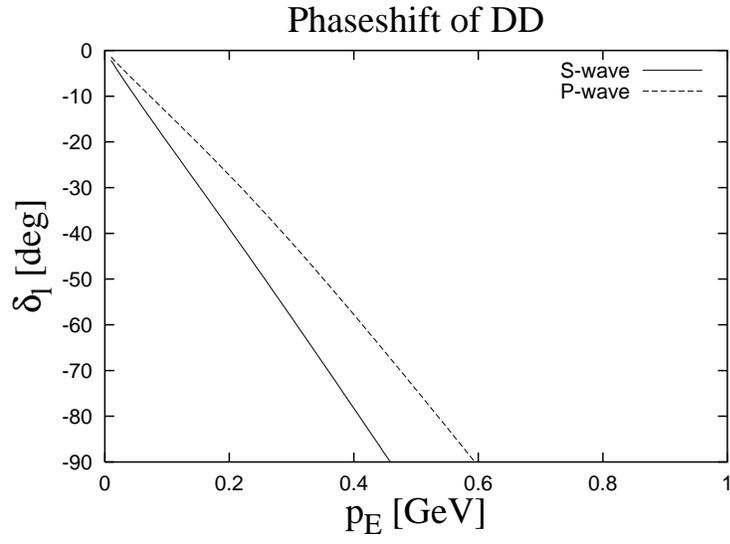,width=10cm}
\end{center}
\caption{Three momentum $p_E$ dependence of the phaseshift
$\delta_l$ for the $DD$ interaction.}
\end{figure}

\begin{figure}[hbtp]
\begin{center}
\epsfig{file=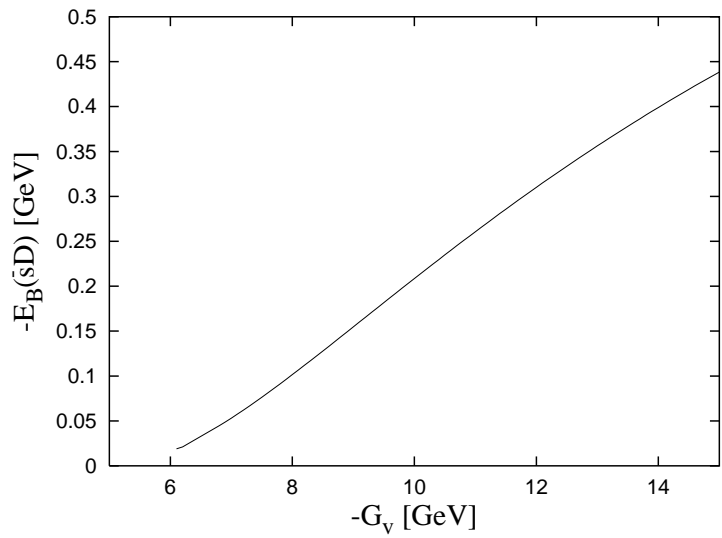,width=10cm}
\end{center}
\caption{$G_v$ dependence of the ${\bar s}$D binding energy.}
\end{figure}



\begin{figure}[hbtp]
\begin{center}
\epsfig{file=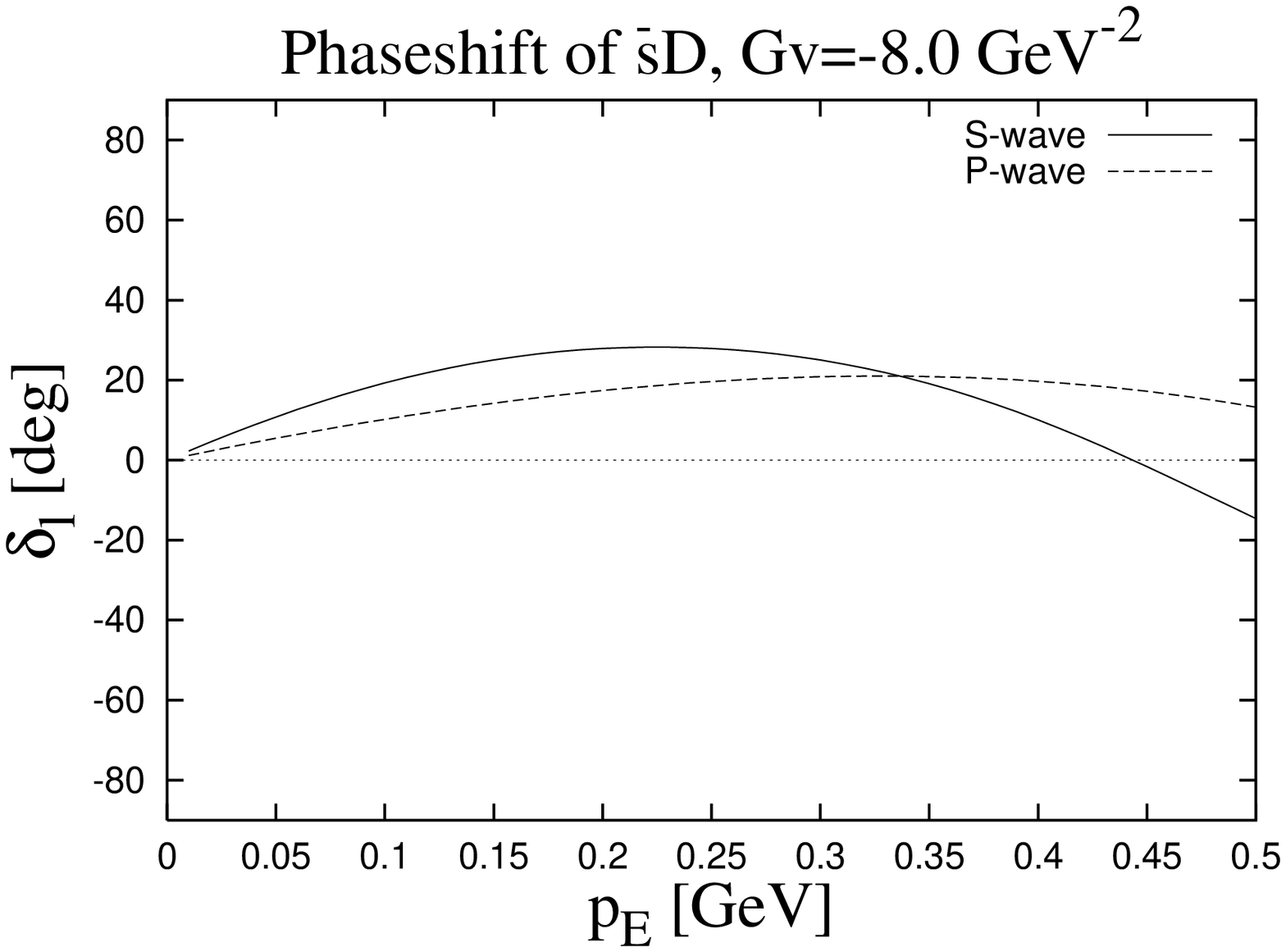,width=10cm}
\end{center}
\caption{Three momentum $p_E$ dependence of the phaseshift
$\delta_l$ for the ${\bar s}D$ interaction
with the coupling constant $G_v=-8.0$ GeV$^{-2}$.}
\end{figure}

\begin{figure}[hbtp]
\begin{center}
\epsfig{file=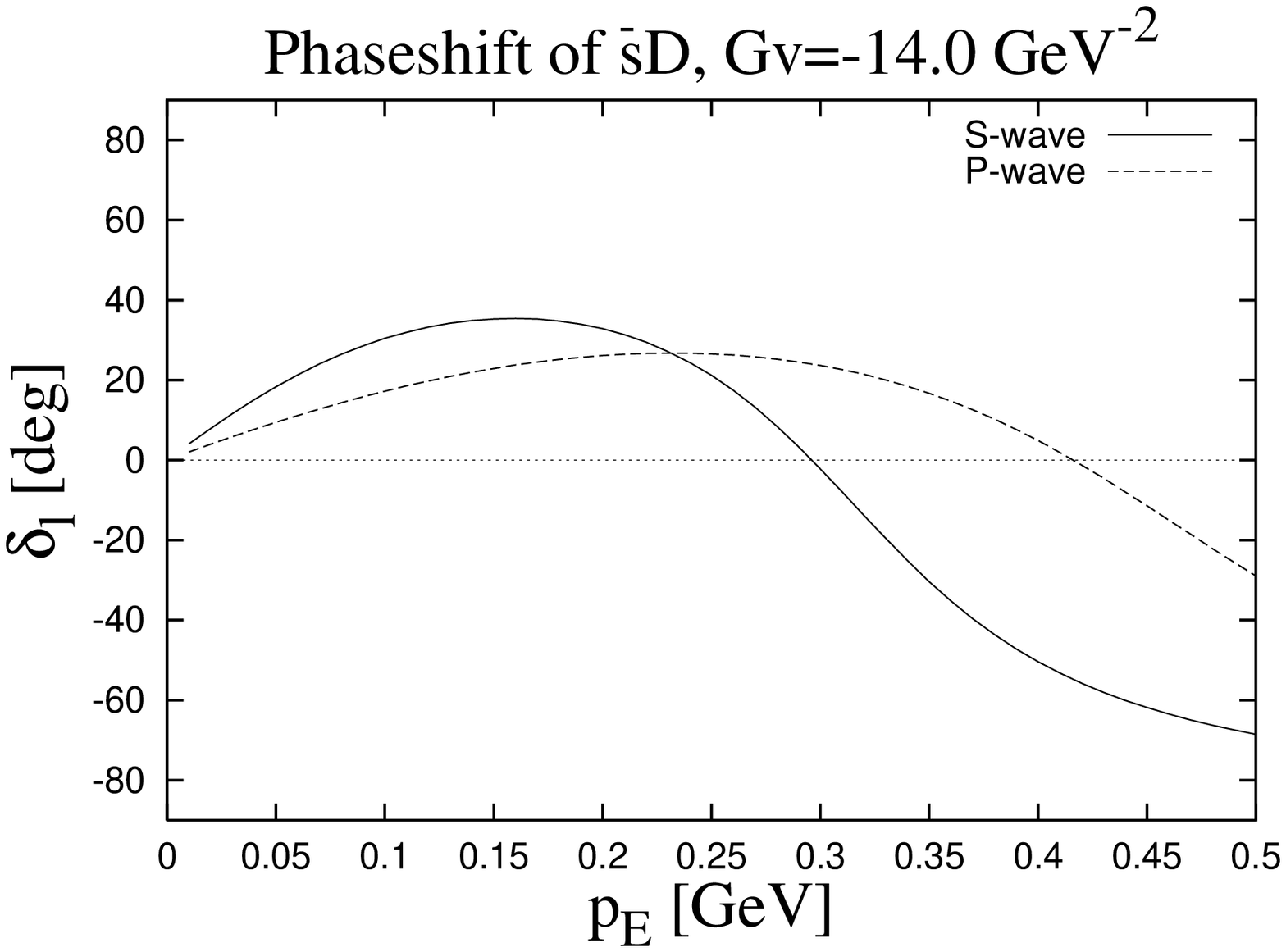,width=10cm}
\end{center}
\caption{Three momentum $p_E$ dependence of the phaseshift
$\delta_l$ for the ${\bar s}D$ interaction
with the coupling constant $G_v=-14.0$ GeV$^{-2}$.}
\end{figure}

\begin{figure}[hbtp]
\begin{center}
\epsfig{file=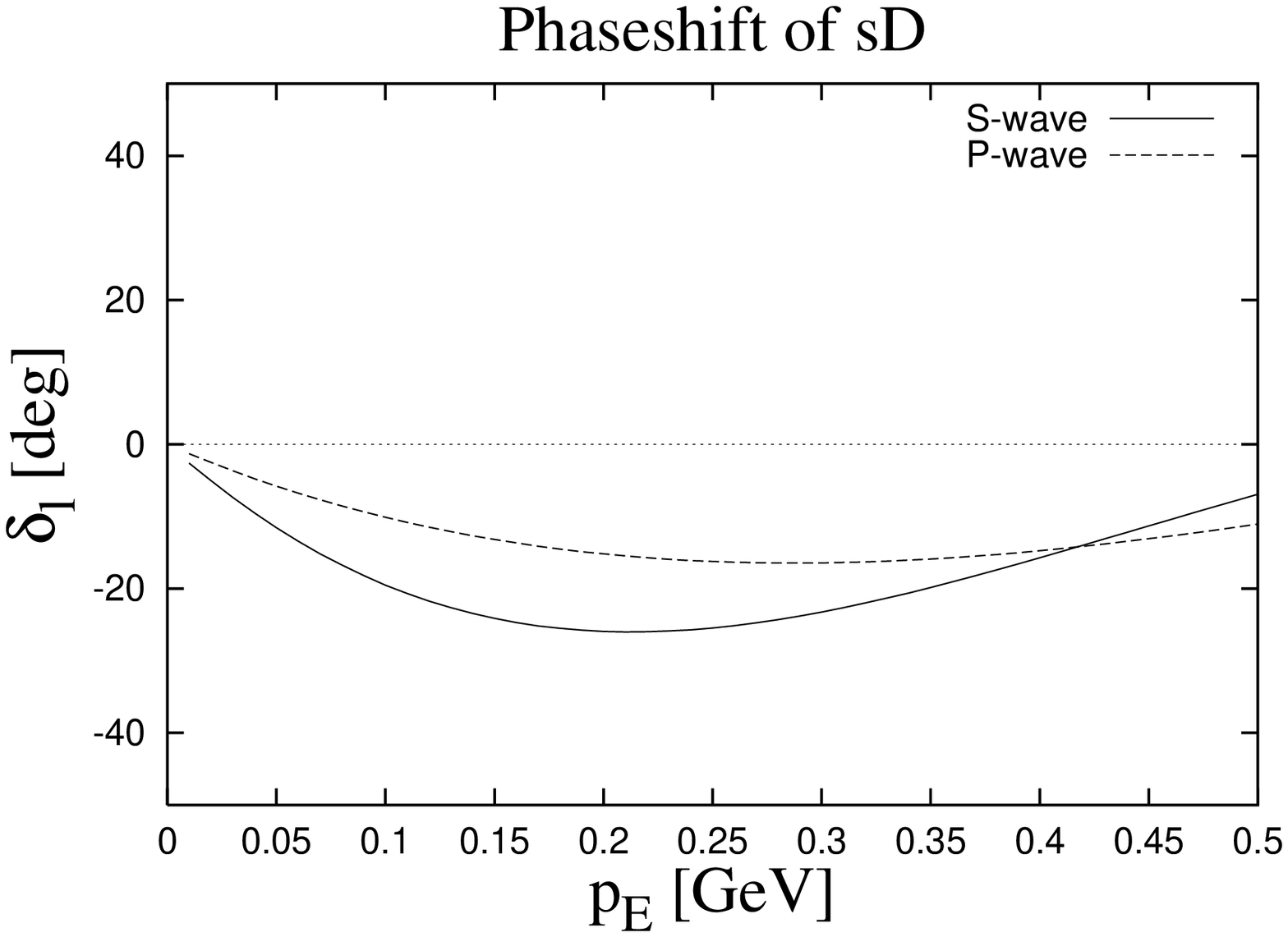,width=10cm}
\end{center}
\caption{Three momentum $p$ dependence of the phaseshift
$\delta_l$ for the $sD$ interaction
with the coupling constant $G_v=-14.0$ GeV$^{-2}$}
\end{figure}

\newpage
\appendix{\LARGE Appendices}

\section{Partial wave expansion}

In the 2-body center of mass frame
the partial wave expansion is defined by
\ba
t(\vec{p}_f,\vec{p}_i) &=&
\sum_l \frac{2l+1}{4\pi} P_l(cos \theta_{p_i p_f })
<p_f l|t|p_i l>\nonumber\\
&\equiv&\sum_l (2l+1) P_l(cos \theta_{p_i p_f }) t^{l}
(|\vec{p}_f|,|\vec{p}_i|), \label{PWE} \ea with $
\vec{p}_{i(f)}\equiv\vec{p\,}_{1i(f)}=-\vec{p\,}_{2i(f)}$. Then
$t^{l}(|\vec{p}_f|,|\vec{p}_i|)$ in Eq. (\ref{PWE}) is written in
terms of $t(\vec{p}_f,\vec{p}_i)$ by \be t^l
(|\vec{p}_f|,|\vec{p}_i|) =\frac12 \int_{-1}^1 dcos\theta_{p_i
p_f} P_{l}(cos\theta_{p_i p_f}) t(\vec{p}_f,\vec{p}_i).
\label{inversepartial} \ee

The phase shift $\delta_l$ is given by
\be t^l (p,p)=-\frac{8\pi\sqrt{s_2}}{p} e^{i\delta_l}sin\delta_l,
\ee where $p\equiv |\vec{p}_{1i}|=|\vec{p}_{2i}|=|\vec{p}_{1f}|
=|\vec{p}_{2f}|$ and $s_2=(p_{1i}+p_{2i})^2=(p_{1f}+p_{2f})^2$.

\section{The results for $\tilde{V}_{{\bar s}(s)D,nm}$ and
$\tilde{K}_{{\bar s}(s)D,nm}$ $(n,m=1,2)$}

In this appendix we show the results for $\tilde{V}_{{\bar
s}(s)D,nm}$ and $\tilde{K}_{{\bar s}(s)D,nm}$ $(n,m=1,2)$ defined
in Eqs. (\ref{sDnm1}-\ref{sbDnm2}):

\ba \tilde{V}_{{\bar s}D,11}(p_{Di},p_{Df},x) &=&
\frac{p_{Di}^0+p_{Df}^0}{2}
,\nonumber\\
\tilde{V}_{{\bar s}D,12}(p_{Di},p_{Df},x)
&=&-\frac{p_{Df}+xp_{Di}}{2} =-\frac{p_{{\bar s}f}+xp_{{\bar
s}i}}{2}
,\nonumber\\
\tilde{V}_{{\bar s}D,21}(p_{Di},p_{Df},x) &=&
\frac{p_{Di}+xp_{Df}}{2} =\frac{p_{{\bar s}i}+xp_{{\bar s}f}}{2}
,\nonumber\\
\tilde{V}_{{\bar s}D,22}(p_{Di},p_{Df},x) &=&
\frac{x}{2}(p_{Di}^0+p_{Df}^0) ,\nonumber \ea and \ba
\tilde{K}_{{\bar s}D,11}(p_{Di},p'_D,x_i)&=&
 \frac12 \left[
(p_{Di}^0+p^{'0}_D) M_s + (\sqrt{s_2}-p^{'0}_D)
(p_D^{'0}+p_{Di}^0) +p_D^{'2}
+x_i p'_D p_{Di}  \right],\nonumber\\
\tilde{K}_{{\bar s}D,12}(p_{Di},p'_D,x_i)&=&
 -\frac12\left[(p'_D+x_ip_{Di})
(M_s-\sqrt{s_2}+p_D^{'0})-p'_D(p_{Di}^0+p^{'0}_{D})
\right],\nonumber\\
\tilde{K}_{{\bar s}D,21}(p_{Di},p'_D,x_i)&=&
\frac12 \left[(p_{Di}+x_ip'_D)(M_s+\sqrt{s_2}-p^{'0}_D)
+x_ip'_D(p_{Di}^0+p_D^{'0})\right],\nonumber\\
\tilde{K}_{{\bar s}D,22}(p_{Di},p'_D,x_i)&=&
-\frac12 \left[x_i M_s(p_{Di}^0+p_D^{'0})-(p_{Di}p'_D
+x_ip^{'2}_D)+x_i(p_D^{'0}-\sqrt{s_2})
(p_{Di}^0+p_D^{'0})
\right],\nonumber
\ea
where
$x\equiv \hat{p}_{Di}\cdot \hat{p}_{Df}$,
$x_i\equiv\hat{p}_{Di}\cdot \hat{p}_D^{\,\,'}$.

$\tilde{V}_{{\bar s}D,nm}$ and
$\tilde{K}_{{\bar s}D,nm}$ are related with
$\tilde{V}_{sD,nm}$ and
$\tilde{K}_{sD,nm}$ by
\ba
\tilde{V}_{{\bar s}D,nm}(p,q,x_{pq})&=&-
\tilde{V}_{sD,nm}(p,q,x_{pq}),\nonumber\\
\tilde{K}_{{\bar s}D,nm}(p,q,x_{pq})&=&-
\tilde{K}_{sD,nm}(p,q,x_{pq}).\nonumber
\ea

\section{Parametrizations for $t_{{\bar s}D}$
and $t_{sD}$}

$t_{{\bar s}D}$ can be parametrized as
\be
t_{{\bar s}D}(p_{Di},p_{Df})
=\sum_{\rho,\rho'=\pm} \Lambda_{\rho}\left[
F_S^{\rho\rho'}+F_T^{\rho\rho'}i\sigma_{\mu\nu}p_{Df}^{\mu}p_{Di}^{\nu}
\right]\Lambda_{\rho'},\nonumber
\label{paramtsbD}
\ee
where $\Lambda_{\pm}=\frac{1\pm \gamma_0}{2}$.
Components of $t_{{\bar s}D}$ is written as
\be
t_{{\bar s}D} (p_{Di},p_{Df}) =
\left(\begin{array}{cc}
F_S^{++}+F_T^{++}i\vec{\sigma}\cdot \vec{n} &
F_T^{+-}\vec{\sigma}\cdot \vec{v}\\
F_T^{-+}\vec{\sigma}\cdot \vec{v} &
F_S^{--}+F_T^{--}i\vec{\sigma}\cdot \vec{n}
\end{array}\right),\nonumber
\ee
where $\vec{n}=\vec{p}_{Df}\times\vec{p}_{Di},
\vec{v}=p_{Df}^0 \vec{p}_{Di}-p_{Di}^0\vec{p}_{Df}$,
and $\pm$ means upper and lower components in the spinor space i.e., $(t_{{\bar s}D})_{\rho,\rho'}=
\Lambda_{\rho}t_{{\bar s}D}\Lambda_{\rho'}$.

The decomposition into upper and lower components
in eq. (\ref{sbDnm1}) for
$t_{{\bar s}D}$ gives
\ba
t_{{\bar s}D,11} (p_{Di},p_{Df}) &=& -F_S^{--},\nonumber\\
t_{{\bar s}D,12} (p_{Di},p_{Df}) &=&
- F_T^{-+}(x p_{Df}^0 p_{Di}-p_{Di}^0 p_{Df}),\nonumber\\
t_{{\bar s}D,21} (p_{Di},p_{Df}) &=&
- F_T^{+-}(p_{Df}^0p_{Di}-xp_{Di}^0p_{Df}),\nonumber\\
t_{{\bar s}D,22} (p_{Di},p_{Df}) &=&
- F_T^{++} p_{Di}p_{Df}(x^2-1).\nonumber
\ea

We can parametrize $t_{{ s}D}$ in the same way
(= eq. (\ref{paramtsbD}))
\be
t_{{s}D}(p_{Df},p_{Di})
=\sum_{\rho,\rho'=\pm} \Lambda_{\rho}\left[
F_S^{\rho\rho'}+F_T^{\rho\rho'}i\sigma_{\mu\nu}p_{Df}^{\mu}p_{Di}^{\nu}
\right]\Lambda_{\rho'},\nonumber
\ee
where $\Lambda_{\pm}=\frac{1\pm \gamma_0}{2}$.

Similar to $t_{{\bar s}D}$
the decomposition into upper and lower components
by eq. (\ref{sDnm1}) gives
\ba
t_{{ s}D,11} (p_{Df},p_{Di}) &=& F_S^{++},\nonumber\\
t_{{ s}D,12} (p_{Df},p_{Di}) &=&
F_T^{+-}(p_{Df}^0 p_{Di}-x p_{Di}^0 p_{Df}),\nonumber\\
t_{{ s}D,21} (p_{Df},p_{Di}) &=&
F_T^{-+}(xp_{Df}^0p_{Di}-p_{Di}^0p_{Df}),\nonumber\\
t_{{ s}D,22} (p_{Df},p_{Di}) &=&
F_T^{--} p_{Di}p_{Df}(x^2-1).\nonumber
\ea

\end{document}